\documentstyle[prl,aps,twocolumn,epsfig]{revtex}
\input epsf.sty

\begin{document}
\draft
\flushbottom
\twocolumn[\hsize\textwidth\columnwidth\hsize\csname@twocolumnfalse\endcsname

\title{
Proximity effect and strong coupling superconductivity in
nanostructures built with an STM}
\author{H. Suderow$^{1}$, E. Bascones$^{2,3}$, A. Izquierdo$^{1}$
, F. Guinea$^{2}$, S. Vieira$^{1}$}
\address{
$^{1}$Laboratorio de Bajas Temperaturas$^{*}$, Departamento de
Fisica de la Materia Condensada, Instituto de Ciencia de Materiales
Nicol\'as Cabrera, Universidad Autonoma de Madrid. E-28049 Madrid.
Spain\\$^{2}$Instituto de Ciencia de Materiales de Madrid$^{*}$,
Consejo Superior de Investigaciones Cientificas, Campus de
Cantoblanco. E-28049 Madrid-Spain \\ $^{3}$Department of Physics,
The University of Texas at Austin, Austin, Texas 78712}
\date{\today}
\maketitle


\begin{abstract}
We present high resolution tunneling spectroscopy data at very low
temperatures on superconducting nanostructures of lead built with
an STM. By applying magnetic fields, superconductivity is
restricted to length scales of the order of the coherence length.
We measure the tunneling conductance and analyze the phonon
structure and the low energy DOS. We demonstrate the influence of
the geometry of the system on the magnetic field dependence of the
tunneling density of states, which is gapless in a large range of
fields. The behavior of the features in the tunneling conductance
associated to phonon modes are explained within current models.
\end{abstract}


]

During the last years there has been an increasing interest on the
understanding of the physical properties of the nanoscopic size
objects. This has been promoted by the rapid development of new
experimental tools that permit a direct access to the realm of the
nanoworld. One of the most widely known is the scanning tunneling
microscope (STM)\cite{ReviewSTM}, which  permits the creation of
metallic nanostructures that can be at the same time imaged and
characterized in situ \cite{Rodrigo94,UR97}. STM operation at low
temperatures has produced new and interesting results, many of them
in the field of superconductivity
\cite{Kirtley90,Hess90,Maggio95,Pan00,Rubio01}.

\begin{figure}
\epsfxsize 6cm
\epsfbox{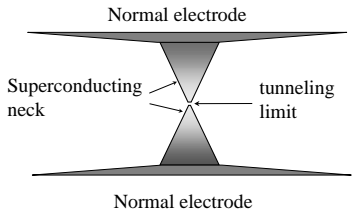}
\caption{
We consider a connecting neck, formed according to the method
described elsewhere
\protect\cite{Rodrigo94,UR97,Suderow00,Suderow00b,Poza98}, broken
into two parts to be able to measure the tunneling density of
states. The diagram is a sketch to show the geometry that we use to
make a simplified model of the nanostructure built with this
procedure.}
\label{fig:Fig1}
\end{figure}

Recently, and using a low temperature STM, we have studied
transport through nanosized metallic
necks\cite{Suderow00,Suderow00b}. In the case of lead we have shown
that it is possible to form a neck of a few hundreds of nanometer
length\cite{Rodrigo94,UR97} which connects the two bulk electrodes
and which can be described as two opposed truncated cones.
Application of an external magnetic field H at temperatures below
the superconducting critical temperature of the bulk creates a very
singular nanostructure when H is higher than bulk critical field,
H$_c$\cite{Poza98}. The bulk electrodes transit to the resistive
state, and superconductivity is confined to the neck. The transport
properties are governed by the diameter of the smallest cross
section, which can be changed in a highly reproducible way from the
large point contact to the single atom point contact
regimes\cite{Suderow00,Suderow00b}. A detailed Ginzburg-Landau
analysis has been done by the authors of \cite{Misko01} to examine
the order parameter in this system.

Here we present new tunneling conductance measurements in these
connecting necks as a function of the magnetic field done by
breaking the neck in-situ into two parts, after its preparation
(described below). This is a singular system with two
superconducting "hills" ending in a sharp tip of atomic dimensions
(fig.1). As the whole procedure is done at low temperatures (about
0.5K), we can neglect atomic diffusion so that the geometry of the
system is not significantly modified when breaking. The conductance
is in the tunneling limit and is proportional to the convolution of
the densities of states (DOS) of both parts of the broken
neck\cite{Wolf}. In this way we gain direct experimental access to
the DOS of a nanoscopic structure with applied and fundamental
interest. Note that it has the same form as two STM tips, one
opposed in front of the other (fig.1). With an in-situ positioning
x-y table we can in principle transport one part of the neck to
another place to use it as a superconducting STM tip that probes a
given sample \cite{SuderowStips}. This is of major interest because
it opens the possibility for new applications as local Josephson
spectroscopy or spin-polarized tunneling, which needs a magnetic
field\cite{Mersevey88,Pan98,Naaman01}. Other methods have been
successfully used to make superconducting STM tips
\cite{Pan98,Naaman01} but, to our knowledge, no data are available
under magnetic fields.

We use a conventional STM set-up in a $^3$He cryostat. The sample
and tip (both of Pb) are cleaned by a mechanical method and mounted
on the STM which is cooled down as fast as possible (about five
minutes) in order to minimize the formation of oxydes on the
surface. The fabrication consists of a series of repeated
indentations done in-situ at low temperatures, as described in
previous work \cite{Rodrigo94,UR97,Suderow00,Suderow00b,Poza98}.
The neck is broken to reach the tunneling regime, where the
estimated work function is of the order of several eV
\cite{Rodrigo94,Suderow01b}, indicating that we have a clean vacuum
tunnel junction between both parts of the neck. The measurement is
done with a tunneling resistance of about $10M\Omega$. Special care
is put on the electronic filtering of the set-up in order to have
the maximal energy (voltage) resolution and to avoid artificial
smearing of the conductance curves. The resolution of the set-up is
of 35$\mu$V, comparable to the lowest temperature of the sample of
400mK.

\begin{figure}
\epsfxsize 6cm
\epsfbox{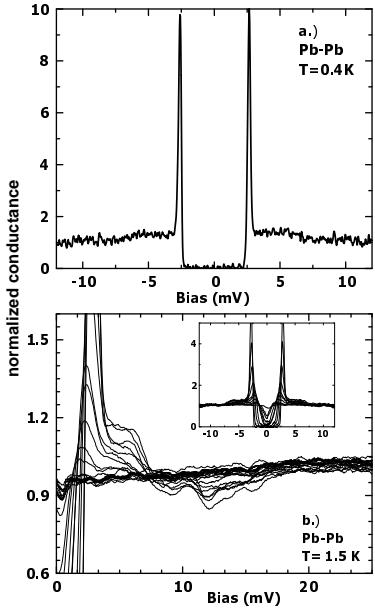}
\caption{
a.)Conductance as a function of the applied bias as measured with
an STM at 0.4K with tip and sample of Pb and zero magnetic field.
b.) The phonon structure as a function of the magnetic field (note
the change in the axis) at 1.5K at fields
0,0.17,0.34,0.51,0.67,0.84,1.01,1.18,1.24 and 1.46 T applied
perpendicularly to the neck. The inset shows the tunneling
conductance in the same units but in a larger range of values.}
\label{fig:Fig1}
\end{figure}

Figure 2a.) shows the conductance at zero field and 0.5K. As Pb is
a strong coupling superconductor, the features due to phonon modes
are clearly observed in the tunneling conductance\cite{SSW}
(fig.2b). According to the well known properties of strong coupling
superconductors, a peak in the effective phonon spectrum
$\alpha^2F(\omega)$ gives a peak in the voltage derivative of the
conductance, located at $\epsilon_{L,T}=2\Delta_0+\omega_{L,T}$ and
therefore the features shown in fig.2b. No significant difference
neither in the value of the superconducting gap $\Delta_0=1.32mV$
nor in the phonon modes ($\omega_{T}=4.4mV$ and $\omega_{L}=8.6mV$)
with respect to planar junction experiments is found within the
experimental resolution ($5\%$)\cite{Wolf}. The magnetic field for
total destruction of the supeconducting correlations depends on the
form of the neck\cite{Suderow01c}. The example shown in the figure
corresponds to a neck having a critical field of about twenty times
the bulk critical field ($H^{Pb}_c(0 K)=0.08 T$). Experiments done
with the field applied parallel or perpendicular to the axis of the
neck show the same behavior, i.e., the destruction of the
superconducting features in the tunneling conductance at magnetic
fields much larger than the bulk critical field, depending on the
form of the neck\cite{Suderow01c}.

To discuss the results, we first try to consider the known example
of thin wires of type I superconductors with lateral dimensions
smaller than the London penetration depth, which remain
superconducting at magnetic fields much higher than the bulk
value\cite{M69}. For instance, superconductivity in an infinite
cylindrical wire can be described by a single pair breaking
parameter given by (in units of $\Delta_0$) $\Gamma_0=\frac{\xi^2
R^2}{3 l_H^4}$, where $\Delta_0$ is the zero field order parameter,
$R$ is the radius of the cylinder and $l_H$ the magnetic length
$l_H=256\AA
/\sqrt{H}$. The peak in the DOS is rounded and
the gap and order parameter are reduced with respect to the zero
field value\cite{M69,Wyder81,Carbotte84,Meyer01}. The gap remains
finite in a large range of fields, and it is only very close to the
critical field (above about $0.95H_c$) where gapless
superconductivity sets in. In the inset of fig.3a we compare the
measured I-V curve with the values of $\Gamma_0$ which best fit the
experiment. Neither the low energy part, nor the magnetic field
dependence are reproduced. The experimental curves show a finite
current at low voltages, corresponding to low energy states within
the gap, even for fields small compared to the critical one. The
density of states at zero voltage is finite already at fields
between 20 and 30 $\%$ the field for complete disappearance of
superconductivity, in clear contradiction with the homogeneous pair
breaking model\cite{M69}.

A better description of the conic, non uniform  geometry of the
necks leads to a more satisfactory result. In a previous work we
proposed\cite{Suderow00}, in the framework of Usadel
equations\cite{U70}, a model in which the field enters as an
effective, position dependent, pair breaking rate. The equations
are solved self-consistently, allowing us to obtain a complete
description of our system in terms of energy and distance to the
center of the broken neck. The main difference with respect to
models using a homogeneous geometry \cite{Poza98,M69} is that there
is a smooth transition to the resistive state as the radius of the
neck increases. The density of states calculated at the center of
the structure remains finite at low energies due to the proximity
effect of the region which is not superconducting, in agreement
with the experiment (fig.3). Using parameters compatible with the
geometry of the measured structure we can get a good fit (fig.3)
for both parallel and perpendicular fields using the same set of
parameters for the whole series of curves measured in a single
structure (see inset of fig.3b, note that not all curves are
presented for clarity).

Previous work has also shown that the conic geometry of our system
needs to be taken into account to understand experiments done in
the single atom point contact limit\cite{Suderow00}, but in that
case, the I-V curves present significant subgap conductance and are
less sensitive to details of the density of states. The tunneling
conductance data however give a straightforward relation between
the I-V curves and the density of states and demonstrate
conclusively that, under field, superconductivity is confined to
the region near the "hills" resulting after the formation of the
neck (fig.1).

\begin{figure}
\epsfxsize 6cm
\centerline{\epsfig{file=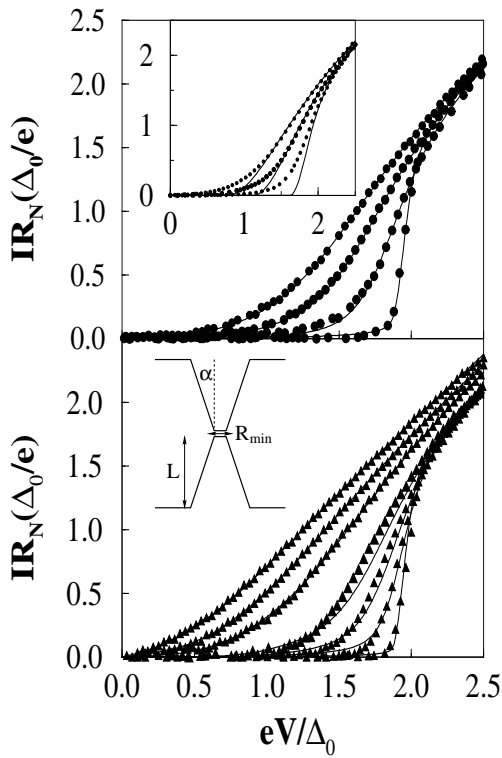,width=2.6in}}
\caption{
Current-voltage characteristics, normalized to the tunneling
resistance and the zero field gap respectively, for different
fields applied parallel at $T=0.4$ K in a) (from bottom to top
0,0.13,0.18 and 0.23 T) and at $T=1.5$ K perpendicular to the axis
of the neck (from bottom to top 0, 0.17, 0.34, 0.84, 1.01, 1.18 T).
Solid lines correspond to the fittings obtained with the geometry
shown in the inset in b). For a) we use $R_{min}=0.8\xi$,
$\alpha=56^{\rm o}$ and $\xi=256$\AA and $L=2.9
\xi$, where $\xi$ is coherence length, and for b), $R_{min}=0.0$,
$\alpha=27^{\rm o}$, $\xi=270$\AA and $L=3.2\xi$. Inset in a) shows
the fittings (lines) to the finite magnetic field experimental
curves in a) in (symbols) using an effective pair breaking
parameter (from bottom to top:$\Gamma_0=0.04,0.13,0.21$).}
\end{figure}

This result demonstrates that it is possible to make STM tips,
which are superconducting even at magnetic fields as high as
several tesla. The proximity effect of the parts of the tip that
transit to the resistive phase at smaller fields needs to be taken
into account in the calculations of their density of states.

\begin{figure}
\epsfxsize 6cm
\epsfbox{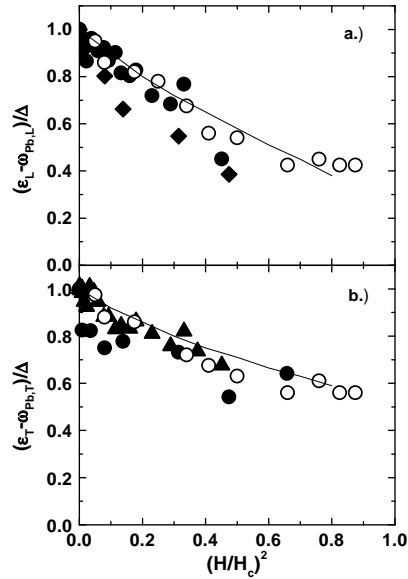}
\caption{
The magnetic field behavior of the energy position $\epsilon_{L,T}$
of the features corresponding to transverse (index T, a.)) and
longitudinal (index L, b.)) phonon energies, for the magnetic field
applied parallel (closed circles) or perpendicular (closed
triangles) to the neck. Superconductivity disappears respectively
at 1.8T and at 0.5 T. Our data are compared to the result obtained
using tunneling spectroscopy in thin films measured with the
magnetic field parallel to the surface(open circles, see
\protect\cite{Wyder81}), and to
the theory of Ref.\protect\cite{Carbotte84} (solid lines). To
determine as accurately as possible the voltage position of the
phonon modes as a function of the magnetic field, we make the mean
value between positive and negative voltages and find
$\epsilon_{L,T}$, the voltage position of the maximum in the second
derivative $d^2I/dV^2$\protect\cite{Wolf}. For $\Delta$ we use two
times the zero field gap value $2\Delta_0$ in our data but
$\Delta_0$ in the data of
\protect\cite{Wyder81}, where an N-S junction was used.
}
\label{fig:Fig6}
\end{figure}

We now discuss the magnetic field behavior of features in the
tunneling conductance associated to the phonon spectrum that appear
at voltages above $2\Delta$ (fig.2b). Note that we have the unique
possibility to follow these features during the confinement of
superconductivity to a system of nanoscopic dimensions. Figure 4
shows the magnetic field behavior of the voltage position of the
features in $dI/dV$ (fig. 2b). We plot
$(\epsilon_{L,T}-\omega_{L,T})$, normalized to $2\Delta_0$, as a
function of the squared magnetic field, normalized to the field for
complete destruction of superconductivity $(H/H_c)^2$, together
with the tunneling spectroscopy measurements in thin films
published in \cite{Wyder81,Carbotte84}. The figure shows the
variation of the voltage position of the phonon modes as a function
of the magnetic field. It does not indicate changes in the phonon
spectrum, but the decrease in the position in energy of the phonon
modes in the superconducting density of states when the pair
breaking effect of the field (and the proximity effect in our case)
destroy the superconducting correlations. The calculation of the
authors of Ref.\cite{Carbotte84} fits the experimental result
within error (solid lines in Fig.4). It even reproduces the
stronger decay in the higher energy longitudinal phonons,
introducing pair breaking effects into the Eliashberg
equations\cite{Carbotte84,Eliashberg}. To make the same approach
using a position dependent pair breaking parameter\cite{Suderow01c}
is a formidable task which would require to solve self-consistently
(both in energy and position) the Eliashberg
equations\cite{Eliashberg}. The figure 4 shows, however, that the
result is the same as in thin films and is compatible with the more
commonly used pair breaking theory\cite{M69}. It demonstrates that
the features related to strong coupling superconductivity are not
sensitive to the precise form of the density of states, and it
confirms previous work \cite{Wyder81,Carbotte84} about strong
coupling superconductivity in the presence of pair breaking. Note
that our experiment gives the additional check that the result is
the same independent of the direction of the magnetic field or the
geometry (data in thin films needed to be done with the magnetic
field applied parallel to the surface).

In summary, we have examined superconducting connecting necks under
magnetic fields and demonstrated the necessity of taking into
account the proximity effect to explain their behavior. We clearly
show that in nanofabricated STM tips made of type I materials,
superconductivity is restricted to a nanoscopic region and remains
at even high magnetic fields. This can considerably extend the
application of STM as a probe. We have also demonstrated that in
this system, the phonon structure of the density of states is not
affected by the size reduction of the superconducting part.

We specially acknowledge discussions with J.T. Devreese, V.M. Fomin
and W. Belzig and support from the ESF programme Vortex Matter in
Superconductors at Extreme Scales and Conditions (VORTEX). We also
acknowledge financial support from the the CICyT (Spain) through
grants DGICYT PB97-0068, the Fermi liquid instabilities ESF
programme (FERLIN) and the Comunidad Aut\'onoma de Madrid. One of
us (E.B.) also thanks financial support from the NSF-DMR0115947 and
the Welch foundation.

$^{*}$ Grupo Intercentros de Bajas Temperaturas, Unidad Asociada
ICMM-CSIC and LBT-UAM.


\begin{references}

\bibitem{ReviewSTM}
See e.g. C.J. Chen, "Introduction to Scanning Tunneling
Microscopy", Oxford Series in Optical and Imaging Sciences, Oxford
University Press (1993).

\bibitem{Rodrigo94}
N. Agra{\"\i}t, J.G. Rodrigo and S. Vieira, Phys. Rev. B, {\bf 48},
8499 (1993); J.G. Rodrigo, N. Agra{\"\i}t and S. Vieira, Phys. Rev.
B, {\bf 50}, 374 (1994).

\bibitem{UR97}
C. Untiedt, G. Rubio, S. Vieira and N. Agra{\"\i}t, Phys. Rev. B
{\bf 56}, 2154 (1997).

\bibitem{Kirtley90}
J.R. Kirtley, J. of Mod. Phys. B, {\bf 4}, 201 (1990).

\bibitem{Hess90}
H.F. Hess, R.B. Robinson, J.V. Waszczak, Phys. Rev. Lett. {\bf 64},
p. 2711 (1990).

\bibitem{Maggio95}
I. Maggio-Aprili, Ch. Renner, A. Erb, E. Walker, O. Fisher, Phys.
Rev. Lett. {\bf 75},  2754 (1995).

\bibitem{Pan00}
S.H. Pan, E.W. Hudson, K.M. Lang, H. Eisaki, S. Uchida, J.C. Davis,
Nature {\bf 403},  746 (2000).

\bibitem{Rubio01}
G. Rubio-Bollinger, H. Suderow, S. Vieira, Physical Review Letters,
{\bf 86},  5582 (2001).

\bibitem{Suderow00}
H. Suderow, E. Bascones, W. Belzig, F. Guinea, S. Vieira
Europhysics Letters, {\bf 50},  749 (2000)

\bibitem{Suderow00b}
H. Suderow, S. Vieira, Physics Letters A, {\bf 275},  299 (2000).

\bibitem{Poza98}
M. Poza, E. Bascones, J. G. Rodrigo, N. Agra{\"\i}t, S. Vieira and
F. Guinea, Phys. Rev. B {\bf 58}, 11173 (1998).

\bibitem{Misko01}
V.R. Misko, V.M. Fomin and J.T. Devreese, Phys. Rev. B, {\bf 64},  14517 (2001).

\bibitem{Wolf}
E.L. Wolf, "Principles of Electron Tunneling Spectroscopy", Oxford
University Press (1989).

\bibitem{SuderowStips}
H. Suderow, S. Vieira, in preparation.

\bibitem{Mersevey88}
R. Mersevey, Phys. Scr. {\bf 38}, 272 (1988).

\bibitem{Pan98}
S.H. Pan, E.W. Hudson, J.C. Davis, Applied Phys. Lett., {\bf 73},
 2992 (1998).

\bibitem{Naaman01}
O. Naaman, W. Teizer, R.C. Dynes, Phys. Rev. Lett., {\bf 87}, 97004
(2001); Rev. Sci. Instrum., {\bf 72}, 1688 (2001).

\bibitem{Suderow01b}
H. Suderow et al., to appear in Physica B.

\bibitem{SSW}
D.J. Scalapino, J.R. Schrieffer and J.W. Wilkins, Phys. Rev., {\bf 148}, 263 (1966).

\bibitem{Suderow01c}
The dependence of the critical field as a function of the form of
the neck will be published in a future work (J.G. Rodrigo, H.
Suderow, S. Vieira, in preparation).

\bibitem{thinfilms}
Provided that the field is applied parallel to the surface of the
sample.

\bibitem{M69}
K. Maki in "Superconductivity", vol. 2, R. D. Parks ed.,
M. Dekker (New York, 1969).

\bibitem{Wyder81}
Th.H.M. Rasing, H. W. M. Salemink, P. Wyder, S. Strassler, Phys.
Rev. B, {\bf 23},  4470 (1981).

\bibitem{Carbotte84}
J.M. Daams, H.G. Zarate, J.P. Carbotte, Phys. Rev. B, {\bf 30},
2577 (1984).

\bibitem{Meyer01}
J.S. Meyer, B.D. Simons, cond-mat/0111039 also discuss pair
breaking in thin films beyond mean field.

\bibitem{U70}
K. D. Usadel, Phys. Rev. Lett. {\bf 25}, 507 (1970).

\bibitem{Eliashberg}
G.M. Eliashberg, Zh. Eksperim. i Teor. Fiz. {\bf 38}, 966
(1960)[Soviet. Phys. JETP {\bf 11}, 696 (1960)].

\end{references}
\end{document}